\documentclass[a4paper,12pt]{article}

\usepackage{amsmath}
\usepackage{amssymb}
\usepackage{amsthm} 
\usepackage[T1]{fontenc}
\usepackage[bf,small,tableposition=top]{caption}
\usepackage{subfig}
\usepackage{setspace}
 \usepackage{multirow}
\usepackage{graphicx}
\usepackage{epstopdf}
\usepackage{array}
\usepackage{lscape}
\usepackage{soul}
\usepackage{bm}
\usepackage{textcomp}
\usepackage{enumitem}
\usepackage{algorithmicx}
\usepackage{lettrine}
\usepackage{tikz}
\usepackage{makeidx}
\usepackage[algoruled,linesnumbered,noline]{algorithm2e}
\usepackage{booktabs,caption,fixltx2e}
\usepackage[flushleft]{threeparttable}
\usepackage{rotating}
\usepackage{natbib}
\usepackage{tikz}
\usetikzlibrary{positioning,decorations.pathreplacing,arrows,shapes,trees}

\theoremstyle{plain}

\begin{document}

\begin{center}
 \Large\bf{ Determination of Bayesian optimal warranty length under Type-II unified hybrid censoring scheme}
\end{center}
 \begin{center}
	\bf {\footnotesize Tanmay Sen$^{1}$, Ritwik Bhattacharya\footnote[2]{\textit{Corresponding author}: Ritwik Bhattacharya (ritwik.bhatta$@$gmail.com)}, Biswabrata Pradhan$^{3}$ and Yogesh Mani Tripathi$^4$}
\end{center}
\begin{center}
	\noindent\textit{\scriptsize $^1$ Data Science and Artificial Intelligence Unit, Ericsson, Kolkata 700135, India }\\
	\noindent\textit{\scriptsize $^2$ Department of Industrial Engineering, School of Engineering and Sciences, Tecnol\'{o}gico de Monterrey, Quer\'{e}taro 76130,  M\'{e}xico}\\
	\noindent\textit{\scriptsize $^{3}$ SQC and OR Unit, Indian Statistical Institute, Kolkata 700108, India} \\
	\noindent\textit{\scriptsize $^4$ Department of Mathematics, Indian Institute of Technology, Patna 801106, India }\\
\end{center}

\begin{abstract}
  Determination of an appropriate warranty length for the lifetime of the product is an important issue to the manufacturer. In this article, optimal warranty length of the product for the combined free replacement and the pro-rata warranty policy is computed based on the Type-II unified hybrid censored data. A non-linear pro-rata warranty policy is proposed in this context. The optimal warranty length is obtained by maximizing an expected utility function. The expectation is taken with respect to the posterior predictive model for the time-to-failure data. It is observed that the non-linear pro-rata warranty policy gives a larger warranty length with maximum profit as compared to linear warranty policy. Finally, a real-data set is analyzed in order to illustrate the advantage of using non-linear pro-rata warranty policy.     
\end{abstract}


\section{Introduction} \label{sec1}
\paragraph{}
Warranty analysis of a manufactured product is an integral part of statistical quality improvement. Improper warranty analysis may affect the business goal and the perceived quality can be in turmoil. Warranty is defined \citep[see][]{blischke_2011} as a contractual agreement between manufacturer (or seller) and consumer (or buyer) that is entered into upon sale of a product. This contract defines the compensation available to the buyer if the performance of the product is found to be unsatisfactory. Therefore, by providing warranties, manufacturer ensures the consumer about the product reliability. Hence, a longer warranty period usually ensures the consumer higher reliability of the product. However, if the product reliability is low, but manufacturer offers an unrealistically large warranty period, then it may incur high penalty cost to the manufacturer. Also, if the warranty period is smaller in comparison with the other competitors in the market, then sales volume of the product may decrease. Therefore, finding an appropriate warranty period is an important task for the manufacturer. The usual way to find the warranty length is based on the assessment of  product reliability. The reliability assessment is typically done through a life-testing experiment. In practice, life-tests are conducted under various censoring schemes in order to save time and cost of the experimentation.\\

\indent In this article, we consider Type-II unified hybrid censoring scheme, abbreviated as Type-II UHCS, \citep[see][]{Balakrishnan_2008} which is the generalization of the generalized Type-I and Type-II hybrid censoring schemes \citep[see][]{Chandrasekar_2004}. The Type-II UHCS can be described as follows. The testing starts with $n$ units and alongside two integers $l, r\in \{1, 2,\cdots, n\}$ and two time points $T_1, T_2 \in (0, \infty)$ are chosen such that $l<r$ and $T_1<T_2$. If the $r$th failure occurs before time $T_1$, terminate the test at $T_1$. If the $l$th failure occurs before $T_1$ and $r$th failure occurs between $T_1$ and $T_2$, terminate the test at $r$th failure time. If the $l$th failure occurs before $T_1$ and $r$th failure occurs after $T_2$, terminate the test at $T_2$. If the $l$th failure occurs after $T_1$ and $r$th failure occurs before $T_2$, terminate the experiment at  $r$th failure time. If the $l$th failure occurs after $T_1$ and $r$th failure occurs after $T_2$, terminate the test at $T_2$. Finally, if the $l$th failure occurs after time $T_2$, terminate the experiment at $l$th failure time. The advantage of Type-II UHCS is that it ensures at least $l$  failures and the maximum test duration is $T_2$. A schematic representation of Type-II UHCS is presented in Figure 1. \\

 The most commonly used warranty policies are free replacement warranty (FRW) policy, pro-rata warranty (PRW) policy and combined FRW/PRW policy \citep[see][]{blischke_2006, blischke_2011}. An important feature of a warranty policy is that if the product fails during the warranty period, consumer will get full or pro-rated compensation from the manufacturer. Under the FRW policy, if the product fails during the warranty period, a non-repairable product is replaced by an identical one free of charge. In case of repairable product, the manufacturer will repair the product free of cost. On the other hand, if the product fails under PRW policy, the manufacturer will provide a pro-rated compensation to the consumer. Sometimes, a combination of both the policies are also considered which is termed as FRW-PRW policy. We consider determination of warranty length for a combined FRW/PRW policy based on data observed under Type-II UHCS. Although there are many works on determination of warranty length for different policies based on complete data  \citep[see][]{menezes1992approach, decroix1999optimal, wu2006determination}, there are few works under censored data. \cite{Christen_2006} determined Bayesian optimal warranty length under pro-rata warranty policy where they considered two-parameter Weibull distribution as product lifetime. \cite{wu_2010} investigated a decision problem under combined FRW-PRW policy. They used a Bayesian approach to determine the optimal warranty lengths under Type-II progressive censoring scheme for a Rayleigh distribution. \cite{chakrabarty2019optimum} investigated optimal reliability acceptance sampling plans under Type-I hybrid censoring schemes by taking warranty cost as constraint. \cite{budhiraja2019optimum} developed optimal reliability acceptance sampling plans under progressive Type-I interval censoring with random removal using a cost model which consists of warranty cost as a component. The aim of this article is two-fold. First, a generalized censoring scheme is used to develop the proposed methodologies, therefore, it can be easily extended to the other censoring schemes which are the special cases of Type-II UHCS. Second, this article proposes a non-linear pro-rated rebate cost and compared it with linear pro-rated rebate cost proposed by \cite{wu_2010}. It has been shown that the proposed non-linear rebate function gives a larger warranty period with maximum profit in comparison with the linear rebate function.\\

The organization of the paper is as follows. In Section 2, lifetime model and posterior predictive distribution based on the data obtained through Type-II UHCS are discussed. In Section 3, we have derived various non-linear cost functions such as rebate function, economic benefit function, warranty cost function, dissatisfaction cost function. Using the cost functions, an utility function is constructed in this section which is maximized to compute optimal warranty length. The computational methodology to obtain optimal warranty lengths is discussed in Section 4. A real-life data are analyzed to illustrate the proposed method in Section 5 and, finally, some concluding remarks are made in Section 6.

\begin{figure}[htbp]
	\tiny
	\begin{tikzpicture}[scale=0.7]
    \draw[thick] (0,0) -- (6,0);
	\draw[thick,dotted] (6,0) -- (8,0);
	\draw[thick] (8,0) -- (18,0);
	\draw [fill] (0, 0) circle [radius=0.1];
	\node[left] at (0,0) {\tiny{\textbf{Start at 0}}};
	\draw [fill] (1, 0) circle [radius=0.1];
	\draw[->] (1,0) -- (2,2);
	\node[below] at (1,0) {\tiny{$X_{1:n}$}};
	\node[right] at (2,2) {\tiny{1st failure}};
	\draw [fill] (4, 0) circle [radius=0.1];
	\draw[->] (4,0) -- (5,2);
	\node[below] at (4,0) {\tiny{$X_{2:n}$}};
	\node[right] at (5,2) {\tiny{2nd failure}};
    \draw [fill] (9, 0) circle [radius=0.1];
	\draw[->](9,0) -- (10,2);
	\node[below] at (9,0) {\tiny{$X_{l:n}$}};
	\node[right] at (10,2) {\tiny{lth failure}};
	\draw [fill] (12, 0) circle [radius=0.1];
	\draw[->](12,0) -- (13,2);
	\node[below] at (12,0) {\tiny{$X_{r:n}$}};
	\node[right] at (13,2) {\tiny{rth failure}};
	\draw [fill] (14, 0) circle [radius=0.1];
	\node[below] at (14,0) {\tiny{$T_1$}};
	\node[below] at (14, -0.5) {\tiny{\textbf{Stop at $T_1$}}};
	\draw [fill] (16, 0) circle [radius=0.1];
	\node[below] at (16,0) {\tiny{$T_2$}};
	\node[right] at (18, 0) {};
	\node at (7, -1) {\scriptsize{\textbf{Case I}}};
	\draw[thick] (0,-5) -- (6,-5);
	\draw[thick,dotted] (6,-5) -- (8,-5);
	\draw[thick] (8,-5) -- (18,-5);
	\draw [fill] (0,-5) circle [radius=0.1];
	\node[left] at (0,-5) {\tiny{\textbf{Start at 0}}};
	\draw [fill] (1,-5) circle [radius=0.1];
	\draw[->] (1,-5) -- (2,-3);
	\node[below] at (1,-5) {\tiny{$X_{1:n}$}};
	\node[right] at (2,-3) {\tiny{1st failure}};
	\draw [fill] (4,-5) circle [radius=0.1];
	\draw[->] (4,-5) -- (5,-3);
	\node[below] at (4,-5) {\tiny{$X_{2:n}$}};
	\node[right] at (5,-3) {\tiny{2nd failure}};
	\draw [fill] (9,-5) circle [radius=0.1];
	\draw[->] (9,-5) -- (10,-3);
	\node[below] at (9,-5) {\tiny{$X_{l:n}$}};
	\node[right] at (10,-3) {\tiny{lth failure}};
	\draw [fill] (12,-5) circle [radius=0.1];
	\node[below] at (12,-5) {\tiny{$T_1$}};
	\draw [fill] (14,-5) circle [radius=0.1];
	\draw[->] (14,-5) -- (15,-3);
	\node[below] at (14,-5) {\tiny{$X_{r:n}$}};
	\node[right] at (15,-3) {\tiny{rth failure}};
	\node[below] at (14, -5.5) {\tiny{\textbf{Stop at  $X_{r:n}$}}};
	\draw [fill] (16,-5) circle [radius=0.1];
	\node[below] at (16,-5) {\tiny{$T_2$}};
	\node[right] at (18, -5) {};
	\node at (7, -6) {\scriptsize{\textbf{Case II}}};
	\draw[thick] (0,-10) -- (6,-10);
	\draw[thick,dotted] (6,-10) -- (8,-10);
	\draw[thick] (8,-10) -- (18,-10);
	\draw [fill] (0,-10) circle [radius=0.1];
	\node[left] at (0,-10) {\tiny{\textbf{Start at 0}}};
	\draw [fill] (1,-10) circle [radius=0.1];
	\draw[->] (1,-10) -- (2,-8);
	\node[below] at (1,-10) {\tiny{$X_{1:n}$}};
	\node[right] at (2,-8) {\tiny{1st failure}};
	\draw [fill] (4,-10) circle [radius=0.1];
	\draw[->] (4,-10) -- (5,-8);
	\node[below] at (4,-10) {\tiny{$X_{2:n}$}};
	\node[right] at (5,-8) {\tiny{2nd failure}};
	\draw [fill] (9,-10) circle [radius=0.1];
	\draw[->] (9,-10) -- (10,-8);
	\node[below] at (9,-10) {\tiny{$X_{l:n}$}};
	\node[right] at (10,-8) {\tiny{lth failure}};
	\draw [fill] (12,-10) circle [radius=0.1];
	\node[below] at (12,-10) {\tiny{$T_1$}};
	\draw [fill] (14,-10) circle [radius=0.1];
	\node[below] at (14,-10) {\tiny{$T_2$}};
	\node[below] at (14, -10.5) {\tiny{\textbf{Stop at  $T_2$}}};
	\draw [fill] (16,-10) circle [radius=0.1];
	\draw[->] (16,-10) -- (17,-8);
	\node[below] at (16,-10) {\tiny{$X_{r:n}$}};
	\node[right] at (17,-8) {\tiny{rth failure}};
	\node[right] at (18, -10) {};
	\node at (7, -11) {\scriptsize{\textbf{Case III}}};
	\end{tikzpicture}
\end{figure}  
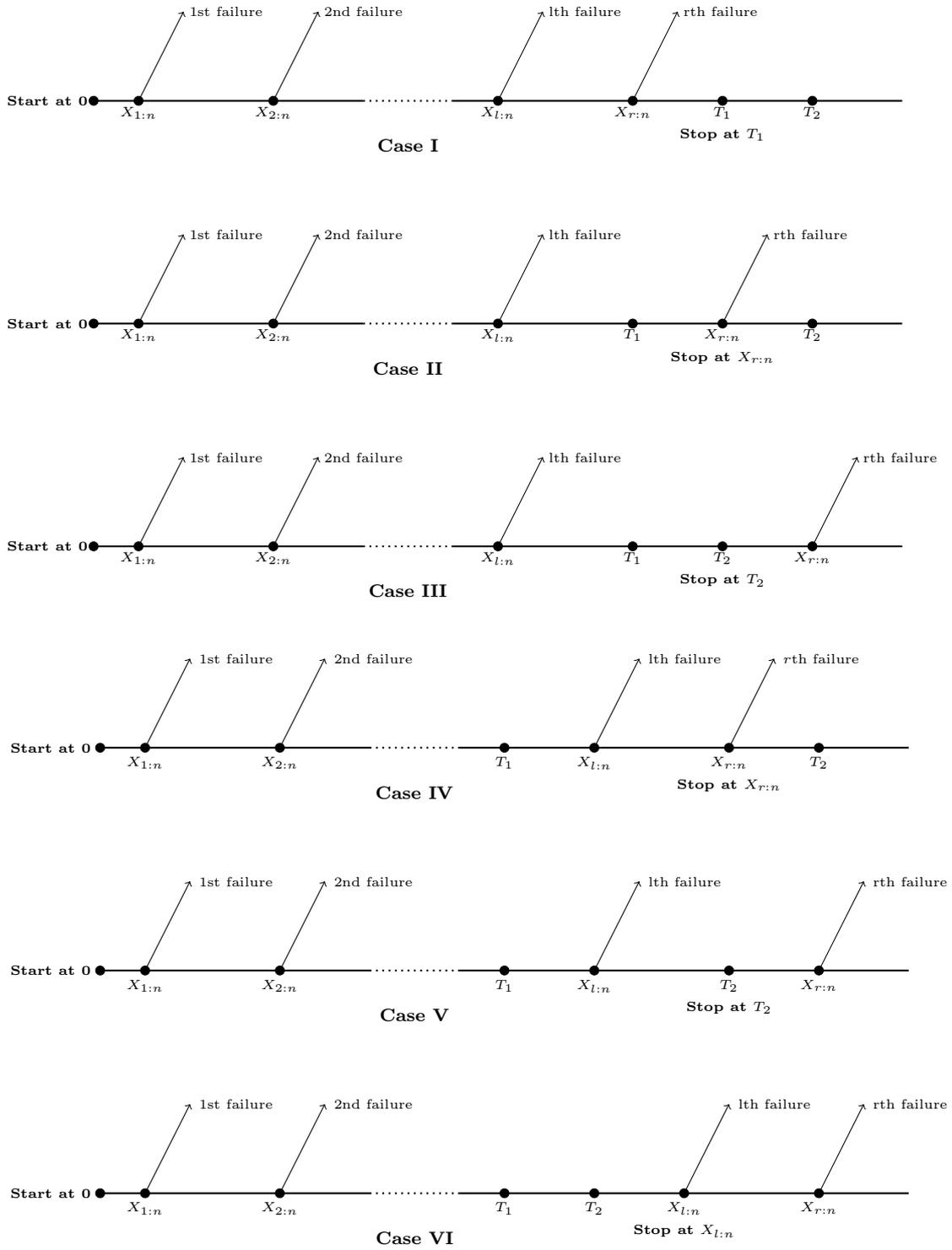
\begin{figure}[htbp]
	\begin{tikzpicture}[scale=0.7]
	
	\draw[thick] (0,0) -- (6,0);
	\draw[thick,dotted] (6,0) -- (8,0);
	\draw[thick] (8,0) -- (18,0);
	
	\draw [fill] (0, 0) circle [radius=0.1];
	\node[left] at (0,0) {\tiny{\textbf{Start at 0}}};
	\draw [fill] (1, 0) circle [radius=0.1];
	\draw[->] (1,0) -- (2,2);
	\node[below] at (1,0) {\tiny{$X_{1:n}$}};
	\node[right] at (2,2) {\tiny{1st failure}};
	
	\draw [fill] (4, 0) circle [radius=0.1];
	\draw[->] (4,0) -- (5,2);
	\node[below] at (4,0) {\tiny{$X_{2:n}$}};
	\node[right] at (5,2) {\tiny{2nd failure}};

	\draw [fill] (9, 0) circle [radius=0.1];
	\node[below] at (9,0) {\tiny{$T_1$}};
	
	\draw [fill] (11, 0) circle [radius=0.1];
	\draw[->](11,0) -- (12,2);
	\node[below] at (11,0) {\tiny{$X_{l:n}$}};
	\node[right] at (12,2) {\tiny{lth failure}};
	
	\draw [fill] (14, 0) circle [radius=0.1];
	\draw[->](14,0) -- (15,2);
	\node[below] at (14,0) {\tiny{$X_{r:n}$}};
	\node[right] at (15,2) {\tiny{$r$th failure}};
	\node[below] at (14, -0.5) {\tiny{\textbf{Stop at $X_{r:n}$}}};

	\draw [fill] (16, 0) circle [radius=0.1];
	\node[below] at (16,0) {\tiny{$T_2$}};
	
	\node[right] at (18, 0) {};
	
	\node at (7, -1) {\scriptsize{\textbf{Case IV}}};
	
	\draw[thick] (0,-5) -- (6,-5);
	\draw[thick,dotted] (6,-5) -- (8,-5);
	\draw[thick] (8,-5) -- (18,-5);
	
	\draw [fill] (0,-5) circle [radius=0.1];
	\node[left] at (0,-5) {\tiny{\textbf{Start at 0}}};
	
	\draw [fill] (1,-5) circle [radius=0.1];
	\draw[->] (1,-5) -- (2,-3);
	\node[below] at (1,-5) {\tiny{$X_{1:n}$}};
	\node[right] at (2,-3) {\tiny{1st failure}};
	
	\draw [fill] (4,-5) circle [radius=0.1];
	\draw[->] (4,-5) -- (5,-3);
	\node[below] at (4,-5) {\tiny{$X_{2:n}$}};
	\node[right] at (5,-3) {\tiny{2nd failure}};
	
	\draw [fill] (9,-5) circle [radius=0.1];
	\node[below] at (9,-5) {\tiny{$T_1$}};
	
	\draw [fill] (11,-5) circle [radius=0.1];
	\draw[->] (11,-5) -- (12,-3);
	\node[below] at (11,-5) {\tiny{$X_{l:n}$}};
	\node[right] at (12,-3) {\tiny{lth failure}};

	\draw [fill] (14,-5) circle [radius=0.1];
	\node[below] at (14,-5) {\tiny{$T_2$}};
	\node[below] at (14, -5.5) {\tiny{\textbf{Stop at  $T_2$}}};
	
	\draw [fill] (16,-5) circle [radius=0.1];
	\draw[->] (16,-5) -- (17,-3);
	\node[below] at (16,-5) {\tiny{$X_{r:n}$}};
	\node[right] at (17,-3) {\tiny{rth failure}};
	\node[right] at (18, -5) {};
	
	\node at (7, -6) {\scriptsize{\textbf{Case V}}};
	
	
	\draw[thick] (0,-10) -- (6,-10);
	\draw[thick,dotted] (6,-10) -- (8,-10);
	\draw[thick] (8,-10) -- (18,-10);
	
	\draw [fill] (0,-10) circle [radius=0.1];
	\node[left] at (0,-10) {\tiny{\textbf{Start at 0}}};
	
	\draw [fill] (1,-10) circle [radius=0.1];
	\draw[->] (1,-10) -- (2,-8);
	\node[below] at (1,-10) {\tiny{$X_{1:n}$}};
	\node[right] at (2,-8) {\tiny{1st failure}};
	
	\draw [fill] (4,-10) circle [radius=0.1];
	\draw[->] (4,-10) -- (5,-8);
	\node[below] at (4,-10) {\tiny{$X_{2:n}$}};
	\node[right] at (5,-8) {\tiny{2nd failure}};
	
	\draw [fill] (9,-10) circle [radius=0.1];
	\node[below] at (9,-10) {\tiny{$T_1$}};
	
	\draw [fill] (11,-10) circle [radius=0.1];
	\node[below] at (11,-10) {\tiny{$T_2$}};

	\draw [fill] (13,-10) circle [radius=0.1];
	\draw[->] (13,-10) -- (14,-8);
	\node[below] at (13,-10) {\tiny{$X_{l:n}$}};
	\node[right] at (14,-8) {\tiny{lth failure}};
	\node[below] at (13, -10.5) {\tiny{\textbf{Stop at  $X_{l:n}$}}};
	
	\draw [fill] (16,-10) circle [radius=0.1];
	\draw[->] (16,-10) -- (17,-8);
	\node[below] at (16,-10) {\tiny{$X_{r:n}$}};
	\node[right] at (17,-8) {\tiny{rth failure}};
	\node[right] at (18, -10) {};
	
	\node at (7, -11) {\scriptsize{\textbf{Case VI}}};
	
	\end{tikzpicture}
	\caption{Schematic representation of Type-II UHCS.}
	\label{UhybFig1}
\end{figure}

\section{Lifetime model and posterior distribution}
\paragraph{}

 Suppose that $X_1, X_2,\cdots, X_n$ are the lifetimes of $n$ testing units which follow a log-normal distribution LN$(\mu, \tau)$. The probability density function (PDF) and the cumulative distribution function (CDF) of LN$(\mu, \tau)$ are given by
\begin{equation}\label{S1E1}
 f_X(x; \mu, \tau) = \sqrt{\frac{\tau}{2 \pi}} x^{-1} e^{-\frac{\tau}{2}(\ln x -\mu)^{2}},\,\, x > 0, \,-\infty < \mu < \infty,\, \tau >0,
\end{equation}
and
\begin{equation}\nonumber
 F_X(x; \mu, \tau) =\Phi \textbf [\sqrt{\tau}\,(\ln x- \mu) \textbf]\,,~~ x>0,
\end{equation} respectively, where $\mu$ and $\tau$ denote unknown parameters of the distribution. Here, $\Phi(\cdot)$ is the CDF of standard normal distribution. The log-normal distribution is quite popular in reliability studies because of the flexibility of its shape \citep[see][]{johnson1994lognormal}. Suppose  $X_{1:n}<X_{2:n}<\cdots<X_{n:n}$ represent corresponding ordered lifetimes. Let $D$ and $\xi$ represent the number of failures and the duration of the life-testing, respectively, under a Type-II UHCS.  Therefore, $(X_{1:n}, X_{2:n},..., X_{D:n}, \xi)$ represents a Type-II UHCS data defined as 
\[ (D,\xi) = \left\{ \begin{array}{lll}
(D_1,T_1) & \mbox{if $X_{l:n}<X_{r:n}<T_1<T_2, ~~ \mbox{where}~~  D_1= r,r+1,\ldots,n$},\\
(r,X_{r:n}) & \mbox{if $X_{l:n}<T_1<X_{r:n}<T_2$},\\
(D_2,T_2) & \mbox{if $X_{l:n}<T_1<T_2<X_{r:n}, ~~ \mbox{where}~~  D_2=l,l+1,\ldots,r-1$},\\
(r,X_{r:n}) & \mbox{if $T_1< X_{l:n}<X_{r:n}<T_2$},\\
(D_2,T_2) & \mbox{if $T_1<X_{l:n}<T_2<X_{r:n}, ~~ \mbox{where}~~  D_2=l,l+1,\ldots, r-1$},\\
(l,X_{l:n}) & \mbox{if $T_1< T_2<X_{l:n}<X_{r:n}$.}
\end{array} \right. \]	
Based on the data obtained by a Type-II UHCS, the likelihood function is given by 
\begin{equation}\label{prolikelihood}
L( \mu, \tau \mid data) \propto \prod_{i=1}^{d} f_X(x_{i:n}; \mu, \tau)\{1-F_X(\xi_0;  \mu, \tau)\}^{n-d},
\end{equation}where, $d$, $\xi_0$ and $x_{i:n}$ are the observed values of $D$, $\xi$ and $X_{i:n}$, respectively. It is  assumed that the joint prior distribution of $(\mu, \tau)$ follows a normal-gamma distribution with probability density function 
\begin{eqnarray}\nonumber
\pi(\mu,\tau | data)&=& \frac{b_1^{a_1}}{\Gamma {a_1}}\sqrt{\frac{q_2}{2\pi}}\,\,\tau^{a_1-\frac{1}{2}}\,\,e^{-\frac{q_2\tau}{2}(\mu-p_2)^2-b_1\tau}\\\nonumber
&=& \frac{b_1^{a_1}}{\Gamma {a_1}} \,\tau^{a_1-1}\,\,e^{-b_1\tau} \times \frac{1}{\sqrt{2\pi}}\frac{1}{\sqrt{\frac{1}{\tau q_2}}}\,e^{-\frac{(\mu-p_2)^2}{2\frac{1}{\tau q_2}}}\\
&=&\pi(\tau) \times \pi(\mu \mid \tau),
\end{eqnarray}
where, $\pi(\tau) \sim Gamma(a_1,b_1)$ and $\pi(\mu \mid \tau) \sim N_{\mu \mid \tau}(p_2, 1/\tau q_2)$. The hyper parameters $a_1,b_1,p_2$ and $q_2$  reflect prior knowledge about unknown parameters of interest, where $a_1,b_1 >0, q_2>0$ and $-\infty<p_2<\infty$. The posterior distribution is given as
\begin{eqnarray}\nonumber
\pi(\mu,\tau| data) = L( \mu, \tau | data) \times \pi(\mu,\tau).
\end{eqnarray}
The posterior predictive distribution, which represents the current beliefs of the decision taker about the failure time, is given as
\begin{eqnarray*}\nonumber
f(t \mid data)  = \int_{-\infty}^{\infty} \int_{0}^{\infty} f(t; \mu, \tau) \pi(\mu,\tau| data) d\mu d\tau.
\end{eqnarray*}

Now we are presenting the expression of the Fisher information $I(\theta)$ about $\theta=(\mu, \tau)$ which will be used in the next section for applying Metropolis-Hastings (MH) algorithm to compute optimal warranty length. The Fisher information $I(\theta)$ under Type-II UHCS is given by \citep[see][]{Sen_2020}
\begin{eqnarray*}	
	I(\theta)=I_{T_1}(\theta)+I_{1,\ldots,l:n}(\theta)+ I_{X_{r:n}\wedge T_2}(\theta) - I_{X_{l:n}\wedge T_2}(\theta)-I_{X_{r:n}\wedge T_1}(\theta),
\end{eqnarray*}
where $I_{T_1}(\theta)$, $I_{1,\ldots,l:n}(\theta)$, and $I_{X_{r:n}\wedge T_2}(\theta)$ represent the Fisher information about $\theta$ under Type-I censoring, Type-II censoring and Type-I hybrid censoring schemes, respectively. The expressions of each of them are given as follows
\begin{eqnarray}	
I_{T_1}(\theta)&=& \int_{0}^{T_1}\bigg \langle \frac{\partial}{\partial \theta}\ln h(x; \theta) \bigg \rangle f_X(x; \theta)\, dx, \\	
I_{1\ldots l:n}(\theta)&=& \int_{0}^{\infty}\bigg \langle \frac{\partial}{\partial \theta}\ln h(x; \theta) \bigg \rangle \sum_{i=1}^{l}f_{i:n}(x; \theta)\, dx,\\
I_{X_{r:n}\wedge T_2}(\theta)&=& \int_{0}^{T_2}\bigg \langle \frac{\partial}{\partial \theta}\ln h(x;\theta) \bigg \rangle \sum_{i=1}^{r}f_{i:n}(x;\theta)\, dx, \label{type1hcs}
\end{eqnarray}
where $h(x; \theta)$ is the hazard function of $X$, $f_{i:n}(x; \theta)$ is the density of $X_{i:n}$, $(\partial/\partial\theta)\ln h(x; \theta)$ is the vector $((\partial/\partial\mu)\ln h(x; \theta), (\partial/\partial\tau)\ln h(x; \theta))^{'}$ and $\langle A \rangle$ is defined as the matrix $A.A^{'},$ for $A\in \mathbb{R}^2$. It may be noted that the expressions of the Fisher information about $\theta$ under Type-I hybrid censored data with schemes $(n,r,T_1)$ and $(n,l,T_2)$ are similar as presented in equation \ref{type1hcs}.

\section{Cost functions}
 
FRW-PRW policy can be viewed as choosing two positive time points $w_1$ and $w_2$ such that $w_1< w_2$, in which FRW policy is applicable in period $[0, w_1)$ and PRW policy is applicable in period $[w_1, w_2)$. Different choices of $w_1$ and $w_2$ raise to FRW or PRW policies as sub-case of  FRW-PRW policy. For instance, when $w_1 = w_2$, it reduces to FRW policy and, when $w_1=0$, it reduces to PRW policy. \cite{wu_2010} considered a linear pro-rata rebate function which is the function of the remaining time of the warranty period. Assuming $S$ be the sales price of a certain product, \cite{wu_2010} defined the cost of reimbursing an item, which is linear in nature under  FRW-PRW policy, as
\[
R_{\tiny{\mbox{cost}}}(t) = 
\begin{cases} 
S, & \text{if }~ 0 \leq t  < w_1, \\
S\, \big(\frac{w_2-t}{w_2-w_1}\big),   & \text{if} ~ w_1 \leq t < w_2,\\
0, & \text{if} ~ t \geq w_2.\\
\end{cases}
\]A pictorial diagram of reimbursing an item under FRW-PRW policy can be visualized in Figure 2.
\begin{figure}[h]
\centering
	\begin{tikzpicture}[scale=4]
	
	\draw [fill] (0, 0) circle [radius=0.01];
	\draw[->] (0,0) -- (2,0) node[right] {$t$}; 
	\draw[->] (0,0) -- (0,0.8)  node[above] {$R_{\tiny{\mbox{cost}}}(t)$};
	\draw[-] (0,0.5) -- (1,0.5);
	\draw[-](1,0)--(1,0.5);
	\draw [fill] (1.8, 0) circle [radius=0.01];
	\draw [fill] (1, 0.5) circle [radius=0.01] node[above] {$(w_1, S)$};
	\draw [fill] (0, 0.5) circle [radius=0.01];
	\draw [fill] (1, 0) circle [radius=0.01];
	\draw[-] (1,0.5) -- (1.8,0);
	\node[below] at (0,0) {{$(0, 0)$}};
	\node[below] at (1,0) {{$(w_1, 0)$}};
	\node[below] at (1.8,0) {{$(w_2, 0)$}};
	\end{tikzpicture}
	\caption{Reimbursing an item under FRW-PRW policy by \cite{wu_2010}}
\end{figure}
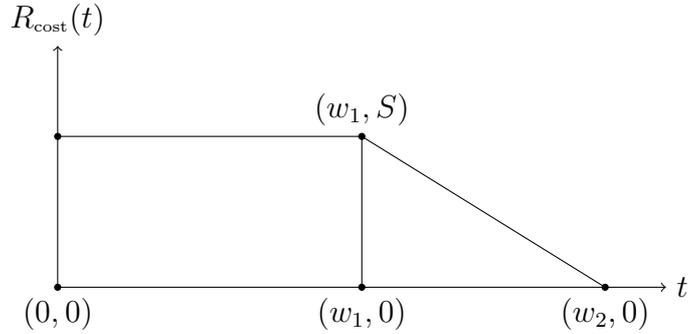In this article, we propose a non-linear rebate function under FRW-PRW policy, which is defined as 
\[
R_{\tiny{\mbox{cost}}}(t) = 
\begin{cases} 
S, & \text{if }~ 0 \leq t  \leq w_1, \\
S\, \Big ( 1-e^{-a\big(\frac{w_2-t}{t-w_1}\big)} \Big),  & \text{if} ~ w_1<t\leq w_2,\\
0, & \text{if} ~ t \geq w_2,\\
\end{cases}
\]where the parameter $a$ ($0<a<1$) controls the non-linearity of the pro-rated rebate function. It is noted that if $a$ increases, then the pro-rated rebate function looks like linear. A graphical representation of this can be seen in Figure 3. Interpretation of Figure 3 is described in Section 5. To compute the optimal warranty length of the products, it is required to construct an utility function, which will be optimized. In this article, three cost functions such as  economic benefit function,  warranty cost function and dissatisfaction cost function are considered which were proposed by \cite{Christen_2006}. In the subsequent sections, we have discussed those cost functions. 
 \begin{figure}[h]
 \centering
   		\includegraphics[width=15cm,height=10cm,angle=0,scale=0.9]{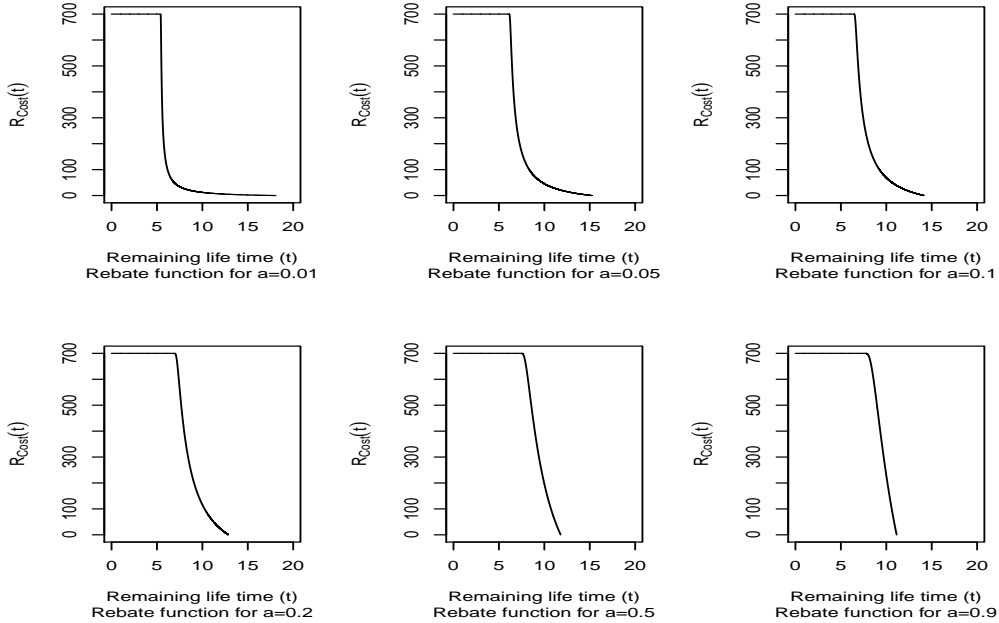}
  		\caption{Non-linear FRW-PRW rebate cost structure}
\end{figure}

\subsection{ Economic benefit function }
\paragraph{}
    By providing a suitable warranty period, the manufacturer may expect an increase in sales volume of the products which results in economic benefit. Therefore, the economic benefit function is considered as the monotone increasing function of the average of two-stage warranty lengths. But, it may be noted that if the manufacturer gives a larger warranty period than their competitor, then the consumer might consolidate certain doubts about the product. As a consequence of that, considering a bounded benefit function should be a realistic choice for the manufacturer. Therefore, an economic benefit function, denoted as $b(w_1,w_2)$, is considered here as \citep[see][]{wu_2010}
  \begin{equation}\nonumber
   b(w_1,w_2)=A_2 M\left(1-e^{-A_1 \left(\frac{w_1+w_2}{2}\right)}\right),
  \end{equation}where $A_2$ is the manufacturers profit for one product, $M$ is the potential number of products to be sold with this warranty policy and  $A_1$ is the parameter to control the speed of increment in benefit. $A_1$ can be uniquely determined from the ratio of two special quantities in the combined FRW/PRW policy. Assuming $t_w$ is the standard market warranty under FRW policy, let us consider the following ratio
  \begin{equation}\nonumber
 \frac{b(0,t_w)}{b(t_w,t_w)} =\frac{1-e^{\frac{A_1 t_w}{2}}}{1-e^{A_1t_w}}.
 \end{equation}The ratio indicates whether the percentage of benefit increases if the manufacturer changes warranty policy from FRW to PRW. Note that $b(w_1,w_2)$ cease to zero when any of the warranty lengths goes to infinity. It interprets the fact that it is unrealistic to expect economic benefit if the warranty length is too long. Let $g(A_1)=(1-e^{\frac{A_1 t_w}{2}})/(1-e^{A_1t_w})$. Then, $g$ is a strictly monotone increasing function, which can take any value between 0.5 and 1. Note that $g(0^+)=0.5$ and $g(\infty)=1$. Therefore, $A_1$ can be determined uniquely by solving the equation $g(A_1)=p^*$ for given $p^*$, where $0.5<p^*<1$. 
  
 \subsection{ Warranty cost function }
 \paragraph{}
  Warranty cost is the direct cost to the manufacturer for reimbursing the products which fail during warranty period. Let us denote warranty cost by $W(t,w_1,w_2)$. It is defined as
  \begin{eqnarray}\nonumber
   W(t,w_1,w_2) &=& \{ \mbox{cost of reimbursing an item $R_{cost}(t)$} \} \times \{ \mbox{the expected}\\\nonumber
   &&\mbox{ number of items that fail under the warranty period} \},
  \end{eqnarray}where $R_{cost}(t)$ is defined in Section 4.1. To find the expected number of failures during the warranty period, we consider the probabilities of failure before time period $w_1$, between time period $w_1$ and $w_2$ and after time period $w_2$. Thus, the warranty cost function can be formulated as 
\begin{eqnarray*}
W(t,w_1,w_2)&=&MF(w_1\mid \mbox{data}) S\,\mathbf{I}_{[0,w_1]}(t)+  \\
&&M\left(F(w_2\mid \mbox{data})-F(w_1\mid \mbox{data})\right)S \left( 1-e^{-a\big(\frac{w_2-t}{t-w_1}\big)} \right)\mathbf{I}_{(w_1,w_2]}(t),
\end{eqnarray*}where $\mathbf{I}_{[,]}(\cdot)$ denotes the indicator function and $F(t|\mbox{data})$ represents the posterior predictive cumulative distribution function.

\subsection{ Dissatisfaction cost function }
\paragraph{}
 We consider another cost function which is the manufacturer's indirect cost to the product. This is called as dissatisfaction cost or penalty cost. Typically, the consumers have certain expectation about the product lifetime. Suppose that the consumer's expected lifetime of the product is $L$, which can be considered as the mean, median or percentile of the posterior predictive distribution. Now, if the product fails during the two-stage warranty period or if it fails immediately after the expiration of combined warranty period, then the consumers have certain dissatisfaction about the product. This can indirectly affect on company's reputation to the buyer and also it can reduce the future sale volumes of the product. Therefore, we split the total dissatisfaction cost into three time intervals as follows.
 \begin{description}
  \item[Case I:] Item fails in the time period $[0,w_1]$,
  \item[Case II:] Item fails in the time period $(w_1,w_2]$,
  \item[Case III:] Item fails in the time period $(w_2, L]$.
 \end{description}
Note that we assume the consumer's expected lifetime $L$ greater than second stage warranty period $w_2$. Customers often seek for higher reliable products and, thus, this assumption is quite practical in nature. \\

 The dissatisfaction cost in Case I (that is, the product fails in the FRW policy) is defined as
\begin{eqnarray}\nonumber
 D_1(t,w_1)&=& \{ \mbox{Proportion~~} q_1~ (0 < q_1 < 1) \mbox{~~of the sales price}\}\times\\\nonumber
 && \{ \mbox{Expected number of failures}\}\\\nonumber
 &=& MF(w_1\mid \mbox{data}) S\,q_1\,\mathbf{I}_{[0,w_1]}(t).
\end{eqnarray}

\noindent Now let us formulate the dissatisfaction cost in case II (that is, the product fails in the PRW policy). We propose a dissatisfaction cost, which is decreased non-linearly with the remaining time of the PRW period. At $w_1$, per unit cost of dissatisfaction is $Sq_1$ and, at $w_2$, per unit cost of dissatisfaction is $Sq_2$ where $0<q_2<1$ with $q_2<q_1$. Therefore, the dissatisfaction cost in Case II is defined as
\begin{eqnarray}\nonumber
 D_2(t,w_1,w_2) &=& M\Big\lbrace F(w_2\mid \mbox{data})-F(w_1\mid \mbox{data})\Big\rbrace \times \\\nonumber
 &&\Big\lbrace Sq_1-(Sq_1-Sq_2) \Big ( 1-e^{-a\big(\frac{w_2-t}{t-w_1}\big)} \Big)\Big\rbrace \mathbf{I}_{(w_1,w_2]}(t).
\end{eqnarray}

\noindent Finally, suppose that the product fails during immediate expiration of warranty and before the consumer's expected lifetime $L(>w_2)$ of the product. Therefore, being an unsatisfied customer, it incurred some dissatisfaction cost. We propose that the
dissatisfaction cost decreases non-linearly with remaining time of consumer's expected lifetime of the product and the dissatisfaction reaches to zero at $L$. Hence, the dissatisfaction cost in Case III is defined as    
\begin{eqnarray*}
 D_3(t,w_1,w_2)&=& M\Big\lbrace F(L\mid \mbox{data})-F(w_2 \mid \mbox{data}) \Big\rbrace \times \\
 && \Big\lbrace Sq_2-Sq_2 \Big ( 1-e^{-a\big(\frac{L-t}{t-w_2}\big)} \Big)\Big\rbrace \mathbf{I}_{(w_2, L]}(t).
\end{eqnarray*}Therefore, the total dissatisfaction cost is given by summing the above three costs and defined as
\begin{equation}\nonumber
 D(t,w_1,w_2)=D_1(t,w_1,w_2)+D_2(t,w_1,w_2)+D_3(t,w_1,w_2).
\end{equation}A pictorial diagram of the dissatisfaction cost structure can be visualized in Figure 4.

 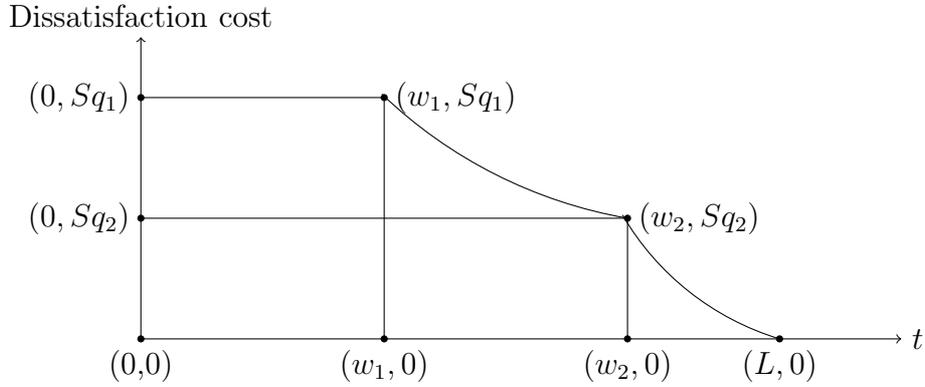
\begin{figure}[h]
	\centering
	\begin{tikzpicture}[scale=4]
	
	\draw [fill] (0, 0) circle [radius=0.01] node[below] {(0,0)};
	\draw[->] (0,0) -- (2.5,0) node[right] {$t$}; 
	\draw[->] (0,0) -- (0,1)  node[above] {Dissatisfaction cost};
	\draw[-] (0,0.8) -- (0.8,0.8);
	\draw[-](0.8,0)--(0.8,0.8);
 	\draw[-](1.6,0)--(1.6,0.4);
 	\draw[-](0,0.4)--(1.6,0.4);
	\draw [fill] (1.6,0.4) circle [radius=0.01] node[right] {$(w_2, Sq_2)$};
	\draw [fill] (1.6, 0) circle [radius=0.01];
	\draw [fill] (0.8,0.8) circle [radius=0.01]node[right] {$(w_1, Sq_1)$} ;
	\draw [fill] (0, 0.4) circle [radius=0.01];
	\draw [fill] (0.8, 0) circle [radius=0.01];
	\draw [fill] (2.1, 0) circle [radius=0.01];
	\draw [fill] (0, 0.8) circle [radius=0.01];
	\node[below] at (0.8, 0) {{$(w_1, 0)$}};
	\node[below] at (1.6, 0) {{$(w_2, 0)$}};
	\node[below] at (2.1, 0) {{$(L, 0)$}};
	\node[left] at (0,0.4) {{$(0, Sq_2)$}};
	\node[left] at (0,0.8) {{$(0, Sq_1)$}};
	\draw[black] (1.6cm,0.4cm)  arc  (-100:-134:1.55);
	\draw[black] (2.1cm,0.0cm)  arc  (-107:-150:0.9);
	\end{tikzpicture}
	\caption{Dissatisfaction cost function structure}
\end{figure}    
    
\subsection{Utility function}
\paragraph{}
Utilizing the three proposed cost functions in Sections 4.2, 4.3 and 4.4, we define an utility function as \citep[see][]{Christen_2006}
\begin{equation}\nonumber
 U(t,w_1,w_2)=b(w_1,w_2)-W(t,w_1,w_2)-D(t,w_1,w_2).
\end{equation}
It is note that the time to failure $t$ is the random quantity in utility function. Therefore, the expected value of the utility function $U(t,w_1,w_2)$ is given by
\begin{eqnarray}
\label{utility}
u^*(w_1,w_2)&=&	E_{\tiny{\mbox{data}}}[U(t,w_1,w_2)| \mbox{data}] \nonumber \\
&=&\int_{0}^{\infty} U(t,{w_1},{w_2}) f(t\mid \mbox{data}) dt \nonumber \\
&= &\int_{-\infty}^{\infty} \int_{0}^{\infty} \bigg\{\int_{0}^{\infty} U(t,{w_1},{w_2}) f(t; \mu, \tau) dt \bigg  \}\pi(\mu\,,\tau)\, d\mu \,\, d\tau \nonumber\\
&=& \int_{-\infty}^{\infty} \int_{0}^{\infty} I_1(\mu\,,\tau)\,\pi(\mu\,,\tau)\, d\mu \,\, d\tau, \label{exputility1}
\end{eqnarray}where
\begin{eqnarray*}
    I_1(\mu\,,\tau)&=&\int_{0}^{\infty} U(t,{w_1},{w_2}) f(t; \mu, \tau) dt \\
     & =& \int_{0}^{\infty} \{ b(w_1,w_2)-W(t,w_1,w_2)-D(t,w_1,w_2) \}f(t; \mu, \tau) dt\\
     & = & b(w_1,w_2)- \int_{0}^{\infty} W(t,w_1,w_2) f(t; \mu, \tau) dt - \int_{0}^{\infty}  D(t,w_1,w_2) \}f(t; \mu, \tau) dt,
\end{eqnarray*}
\begin{eqnarray}
\int_{0}^{\infty} W(t,w_1,w_2) f(t; \mu, \tau) dt &=& S [F(w_1\mid \mbox{data})]^2
+ S[F(w_2 \mid \mbox{data})-F(w_1 \mid \mbox{data})] \nonumber\\ 
&& \times \int_{w_1}^{w_2} \Big ( 1-e^{-a\big(\frac{w_2-t}{t-w_1}\big)} \Big) f(t; \mu, \tau) dt, \nonumber\\
\int_{0}^{\infty} D(t,w_1,w_2) f(t; \mu, \tau) dt &=&\int_{0}^{\infty} \{ D_1(t,w_1)+D_2(t,w_1,w_2)+D_3(t,w_2)\}f(t; \mu, \tau) dt \nonumber\\
&=& S [F(w_1\mid \mbox{data})]^2 +  S[F(w_2 \mid \mbox{data})-F(w_1 \mid \mbox{data})] \nonumber\\\nonumber 
&& \times \int_{w_1}^{w_2} \Big ( 1-e^{-a\big(\frac{w_2-t}{t-w_1}\big)} \Big) f(t; \mu, \tau) dt \\\nonumber
&&+  S[F(L\mid \mbox{data})-F(w_2 \mid \mbox{data})] \nonumber\\ && \times \int_{w_2}^{L} \bigg\{ q_2-q_2\Big ( 1-e^{-a\big(\frac{L-t}{t-w_2}\big)} \Big)\bigg\} f(t; \mu, \tau) dt. \nonumber
\end{eqnarray}    
    
\section{Optimal warranty length}

In this section, we proposed a method to compute optimal warranty length. In order to compute the optimal warranty length, that is, the optimal values of $w_1$ and $w_2$, the expected utility function $u^*(w_1,w_2)$ given in equation \ref{utility} is maximized with respect to $w_1$ and $w_2$. Therefore, the optimal warranty $(w^{*}_1, w^{*}_2)$ is the solution of the following  optimization problem
\begin{equation}\label{exputility2}
 (w^{*}_1, w^{*}_2) = \mbox{arg}\left( \underset{w_1< w_2; w_i\in R^{+}, i=1, 2}{\text{Maximize}}~ u^*(w_1,w_2) \right),
\end{equation}where $R^{+}$ is the set of all positive real numbers. In general, the optimization problem in (\ref{exputility2}) does not have a closed form analytical solution. Nevertheless, in this article, we propose to use Metropolis-Hasting (MH) algorithm to compute the Bayes estimates of (\ref{exputility1}) and, then, using that estimate in (\ref{exputility2}), we compute the optimal solution. The procedure of MH algorithm suggests that the samples from a posterior distribution can be generated using some proposal density. Commonly, a symmetric type proposal density such as $J(({\mu}^*, {\tau}^*) \mid (\mu, \tau) ) = J((\mu, \tau) \mid ({\mu}^*, {\tau}^*))$ can be taken into consideration. Here, we consider a bivariate normal ${N}_{2}(( \mu, \ln \tau)$, $I^{-1}(\mu, \tau))$  proposal density where $I^{-1}(\mu, \tau)$ denotes the inverse of the information matrix. Since we are generating samples from a bivariate normal distribution, few negative observations for $\tau$ may appear, which is not acceptable. In this regard, we propose the following steps of the MH algorithm to draw samples from the corresponding posterior density.

\begin{description}
 \item[Step 1:] Set initial value of $(\mu, \tau)$ as $(\mu, \tau)=(\mu_0, \tau_0)$
 \item[Step 2:] For $i =1, 2, \ldots, N$ repeat the following steps
 \begin{enumerate}[label=(\alph*)]
\item Set $(\mu, \tau) = (\mu_{i-1}, \tau_{i-1})$
\item Generate a new candidate parameter value $\delta$ from $N_{2}((\mu, \ln \tau), I^{-1}(\mu, \tau))$
\item Set $({\mu}^*, {\tau}^*) = (\delta_1, \mbox{exp}(\delta_2))$
\item Calculate ${a}^* = \min\Big( 1, \frac{\pi({\mu}^*, {\tau}^* \mid \bm{x}){\mu}^* {\tau}^*}{\pi(\mu, \tau \mid \bm{x})\mu \tau} \Big)$
\item Update $(\mu_{i}, \tau_{i}) = ({\mu}^*, {\tau}^*)$ with probability ${a}^*$; otherwise set $(\mu_{i}, \tau_{i})=(\mu, \tau)$
\end{enumerate}
\end{description}
The above procedure will generate $N$ observations of $(\mu, \tau)$. Some initial observations of size $N_0$, say, are discarded as burn-in observations and the remaining observations $N-N_0~(= k, \mbox{~say})$ can be used to compute the Bayes estimate of $u^*(w_1,w_2)$ in (\ref{exputility1}). Subsequently, the corresponding Bayes estimate  can be computed as
\begin{equation}\nonumber
 u^*({w_1},{w_2}) =\frac{1}{k}\sum_{i=1}^{k}I_1{(\mu_i,\tau_i)}.
\end{equation}
Finally, the optimal warranty period $(w_1^*, w_2^*)$ in (\ref{exputility2}) is computed by solving the optimization problem
\begin{equation}\nonumber
 \underset{w_1< w_2; w_i\in R^{+}, i=1, 2}{\text{Maximize}}~ \frac{1}{k}\sum_{i=1}^{k}I_1{(\mu_i,\tau_i)}.
\end{equation}
This is a non-linear optimization problem with two real continuous decision variables. Newton-Raphson method can be used to solve this problem. 

\section{Numerical illustration with real-life data analysis}
\paragraph{}
In this section, a real-life data set is considered for illustrative purpose. The data set is taken from \citet{Proschan_1963}. The data represent the intervals of successive failure times in hours of the air conditioning system of  Boeing 7912 jet airplane. The corresponding ordered times are listed as 1, 3, 5, 7, 11, 11, 11, 12, 14, 14, 14, 16, 16, 20, 21, 23, 42, 47, 52, 62, 71, 71, 87, 90, 95, 120, 120, 225, 246, 261.  \citet{Christen_2005} fitted the data with log-normal distribution and calculated the hyper parameter values as 
$a_1=36.9, b_1=29.1, p_2=3.3$ and $q_2=287.9$. By using the censoring scheme $n=30, r=20, l=7, T_1=100, T_2=120$, we have generated Type-II UHCS data as 1, 3, 5, 7, 11, 11, 11, 12, 14, 14, 14, 16, 16, 20, 21, 23, 42, 47, 52, 62, 71, 71, 87, 90, 95. Suppose that the sales price of this product is $S = \$ 700$ and the production cost of the product is $C=\$ 500$. Therefore, the profit per unit product is $A_2 = \$ 200$. The manufacturer gives a standard warranty, which is the $0.1$th quantile of the posterior predictive distribution under the FRW policy i.e. standard warranty is $t_w = 7.245$ hours. Since, we are considering combined FRW/PRW policy, manufacturer is interested to change the warranty policy from FRW to PRW and assumed that the percentage of benefit remains to be $p^*= 0.75$ i.e., $g(A_1)=0.75$. In this case, the unique solution to the  equation $g(A_1)=0.75$  is $A_1=0.303$. Since the customer dissatisfaction indirectly affect on the image of the company, so, the product sales may be reduced. Therefore, we assume that proportions of  customer dissatisfaction are $(q_1, q_2)$ = (0.09, 0.04). The consumer will satisfy with the product if its lifetime reaches customer's expectation over the product lifetime.  Here, we assume customer's expected lifetime of the product is $L=27.263$, which is the median of the posterior predictive distribution. We observed that for linear rebate function, the optimal warranty length under combined FRW/PRW policy is $(w_1^*,w_2^*)=(7.317, 11.641)$ and the maximum value is \$ 175.281M. Using the proposed non-linear rebate function defined in Section 4.1, the optimal warranty lengths under the combined FRW/PRW policy are computed as
\begin{itemize}
 \item For $a=0.01$, $(w_1^*,w_2^*)=(5.388, 18.060)$ and optimal value is \$ 183.622M,
 \item For, $a=0.05$, $(w_1^*,w_2^*)=(6.091, 15.267)$ and optimal value is \$ 180.159M
 \item For $a=0.09$, $(w_1^*,w_2^*)=(6.385, 14.142)$ and optimal value is \$ 178.761M,
 \item For $a=0.2$, $(w_1^*,w_2^*)=(6.839, 12.860)$ and optimal value is \$ 177.013M,
 \item For $a=0.5$, $(w_1^*,w_2^*)=(7.270, 11.796)$ and optimal value is \$ 175.679M,
 \item For $a=0.9$, $(w_1^*,w_2^*)=(7.314, 11.653)$ and optimal value is \$ 175.384M.
\end{itemize}

It is observed that the optimal warranty lengths under non-linear rebate function are wider than that of linear rebate function. Also, optimal warranty length and optimal value  both decrease with increasing $a$. For $a=0.01, 0.05, 0.09, 0.2, 0.5$ and $0.9$, the corresponding non-linear rebate functions are plotted in Figure 3. From Figure 3, it is observed that when $a$ increases, the non-linear pro-rated rebate function looks like a linear pro-rated rebate function.

\section{Conclusion}
\paragraph{}

In our study, We have considered log-normal as the lifetime distribution of the product, however, the proposed methodologies can be extended easily to other lifetime distributions. This article also proposes a non-linear pro-rated rebate cost and compared it with linear pro-rated rebate cost proposed by \cite{wu_2010}. For non-linear pro-rated rebate cost, a larger warranty period with maximum profit is obtained in comparison with the linear rebate cost. This is the prime advantage of choosing a non-linear pro-rated rebate cost function.



\bibliographystyle{apalike}
\bibliography{BaysWarranty}

\begin{thebibliography}{}

\bibitem[Balakrishnan et~al., 2008]{Balakrishnan_2008}
Balakrishnan, N., Rasouli, A., and Farsipour, N.~S. (2008).
\newblock Exact likelihood inference based on an unified hybrid censored sample
  from the exponential distribution.
\newblock {\em Journal of Statistical Computation and Simulation},
  {\bf{78}}:475 -- 488.

\bibitem[Blischke et~al., 2011]{blischke_2011}
Blischke, W.~R., Karim, M.~R., and Murthy, D. N.~P. (2011).
\newblock {\em Warranty data collection and analysis}.
\newblock Springer Science \& Business Media.

\bibitem[Budhiraja and Pradhan, 2019]{budhiraja2019optimum}
Budhiraja, S. and Pradhan, B. (2019).
\newblock Optimum reliability acceptance sampling plans under progressive
  type-i interval censoring with random removal using a cost model.
\newblock {\em Journal of Applied Statistics}, 46(8):1492--1517.

\bibitem[Chakrabarty et~al., 2019]{chakrabarty2019optimum}
Chakrabarty, J.~B., Chowdhury, S., and Roy, S. (2019).
\newblock Optimum life test plan for type-i hybrid censored weibull distributed
  products sold under general rebate warranty.
\newblock {\em International Journal of Production Research}, pages 1--14.

\bibitem[Chandrasekar et~al., 2004]{Chandrasekar_2004}
Chandrasekar, B., Childs, A., and Balakrishnan, N. (2004).
\newblock Exact likelihood inference for the exponential distribution under
  generalized \mbox{Type-I and Type-II} hybrid censoring.
\newblock {\em Naval Research Logistics}, {\bf{51}}:994 -- 1004.

\bibitem[DeCroix, 1999]{decroix1999optimal}
DeCroix, G.~A. (1999).
\newblock Optimal warranties, reliabilities and prices for durable goods in an
  oligopoly.
\newblock {\em European Journal of Operational Research}, 112(3):554--569.

\bibitem[Guti{\'e}rrez-Pulido et~al., 2005]{Christen_2005}
Guti{\'e}rrez-Pulido, H., Aguirre-Torres, V., and Christen, J.~A. (2005).
\newblock A practical method for obtaining prior distributions in reliability.
\newblock {\em IEEE Transactions on Reliability}, {\bf{54}}:262 -- 269.

\bibitem[Guti{\'e}rrez-Pulido et~al., 2006]{Christen_2006}
Guti{\'e}rrez-Pulido, H., Aguirre-Torres, V., and Christen, J.~A. (2006).
\newblock A bayesian approach for the determination of warranty length.
\newblock {\em Journal of Quality Technology}, {\bf{38}}:180 -- 189.

\bibitem[Johnson et~al., 1994]{johnson1994lognormal}
Johnson, N.~L., Kotz, S., and Balakrishnan, N. (1994).
\newblock {\em Continuous univariate distributions Vol 1: Models and
  applications}.
\newblock John Wiley \& Sons, New York.

\bibitem[Menezes and Currim, 1992]{menezes1992approach}
Menezes, M.~A. and Currim, I.~S. (1992).
\newblock An approach for determination of warranty length.
\newblock {\em International Journal of Research in Marketing}, 9(2):177--195.

\bibitem[Murthy and Blischke, 2006]{blischke_2006}
Murthy, D. N.~P. and Blischke, W.~R. (2006).
\newblock {\em Warranty management and product manufacture}.
\newblock Springer Science \& Business Media.

\bibitem[Proschan, 1963]{Proschan_1963}
Proschan, F. (1963).
\newblock Theoretical explanation of observed decreasing failure rate.
\newblock {\em Technometrics}, {\bf{5}}:375 -- 383.

\bibitem[Sen et~al., 2020]{Sen_2020}
Sen, T., Bhattacharya, R., Pradhan, B., and Tripathi, Y.~M. (2020).
\newblock Statistic al inferenc e and \mbox{Bayesian} optimal life-testing
  plans under \mbox{Type-II} unified hybrid censoring sc heme.
\newblock {\em http://arxiv.org/abs/2004.05308}.

\bibitem[Wu et~al., 2006]{wu2006determination}
Wu, C.-C., Lin, P.-C., and Chou, C.-Y. (2006).
\newblock Determination of price and warranty length for a normal lifetime
  distributed product.
\newblock {\em International Journal of Production Economics}, 102(1):95--107.

\bibitem[Wu and Huang, 2010]{wu_2010}
Wu, S.~J. and Huang, S.~R. (2010).
\newblock Optimal warranty length for a rayleigh distributed product with
  progressive censoring.
\newblock {\em IEEE Transactions on Reliability}, {\bf{59}}:661 -- 666.

\end{thebibliography}

\end{document}